\documentclass[usenatbib]{mn2e}
\usepackage{graphicx}

\usepackage{amssymb}
\usepackage{color}
\usepackage{float}

\newcommand{\bn}{\begin{enumerate}}
\newcommand{\en}{\end{enumerate}}
\newcommand{\bi}{\begin{itemize}}
\newcommand{\ei}{\end{itemize}}

\def\gtorder{\mathrel{\raise.3ex\hbox{$>$}\mkern-14mu
    \lower0.6ex\hbox{$\sim$}}}
\def\ltorder{\mathrel{\raise.3ex\hbox{$<$}\mkern-14mu
    \lower0.6ex\hbox{$\sim$}}}

\newcommand{\apj}{ApJ}
\newcommand{\aap}{A\&A}
\newcommand{\apjl}{ApJL}

\newcommand{\mnras}{MNRAS}
\newcommand{\aj}{AJ}

\title[Stellar Bars in Counter-Rotating Dark Matter Halos]{Stellar Bars in Counter-Rotating Dark Matter Halos:\\
The Role of Halo Orbit Reversals} 
\author[Angela Collier, Isaac Shlosman, and Clayton Heller]
{Angela Collier$^{1}$\thanks{E-mail: angela.collier@uky.edu},
Isaac Shlosman$^{1,2}$\thanks{E-mail: shlosman@pa.uky.edu},
Clayton Heller$^{3}$
\\
\footnotemark    
$^{1}$ Department of Physics \& Astronomy, University of Kentucky, Lexington, KY 40506-0055, USA\\
$^{2}$ Theoretical Astrophysics, Graduate School of Science, Osaka University, Osaka 560-0043, Japan\\  
$^{3}$ Department of Physics \& Astronomy, Georgia Southern University, Statesboro, GA 30460, USA \\
}

\begin{document}

\date{Accepted ?; Received ??; in original form ???}


\maketitle

\begin{abstract}
Disk galaxies can exchange angular momentum and baryons with their host dark matter (DM) halos. These halos possess internal spin, $\lambda$, which is insignificant rotationally but does affect interactions between the baryonic and DM components. While statistics of prograde and retrograde spinning halos in galaxies is not available at present, the existence of such halos is important for galaxy evolution. In the previous works, we analyzed dynamical and secular evolution of stellar bars in prograde spinning halos and the DM response to the bar perturbation, and found that it is modified by the resonant interactions between the bar and the DM halo orbits. In the present work, we follow the evolution of stellar bars in retrograde halos. We find, that this evolution differs substantially from evolution in rigid unresponsive halos, discussed in the literature. First, we confirm that the bar instability is delayed progressively along the retrograde $\lambda$ sequence. Second, the bar evolution in the retrograde halos differs also from that in the prograde halos, in that the bars continue to grow substantially over the simulation time of 10\,Gyr. The DM response is also substantially weaker compared to this response in the prograde halos. Third, using orbital spectral analysis of the DM orbital structure, we find a phenomenon we call the orbit reversal --- when retrograde DM orbits interact with the stellar bar, reverse their streaming and precession, and become prograde. This process dominates the inner halo region adjacent  to the bar and allows these orbits to be trapped by the bar, thus increasing efficiency of angular momentum transfer by the Inner Lindblad Resonance. We demonstrate this reversal process explicitly in a number of examples.  
\end{abstract}



\section{Introduction}
\label{sec:intro}

In the current paradigm of galactic structure, the baryonic component is deeply embedded in the massive dark matter (DM) halos. Numerical simulations of DM structure formation in the universe has shown that halos exhibit internal spins \citep[e.g.,][]{peeb69}. The spin parameter can be defined as $\lambda\equiv J/J_\textrm{max}$, where $J$ is the halo's angular momentum, $J_\textrm{max}$ is its Keplerian maximum, and its range is practically limited to $\lambda\sim 0-0.1$ \citep[e.g.,][]{bull01}. The halo spin distribution can be fit by a lognormal distribution,

\begin{equation}
P(\lambda) = \frac{1}{\lambda (2\pi\sigma)^{1/2}} \textrm{exp} \bigg[-\frac{\textrm{ln}^2 (\lambda/\lambda_0)}{2\sigma^2}\bigg],
\end{equation}
where $\lambda_0=0.035\pm 0.005$ and $\sigma=0.5\pm 0.03$ are the fitting parameters \citep[][]{bull01}. While the spin is insignificant rotationally, the majority of DM halos must be spinning to some extent.

Owing to a complex assembly history of individual halos, their angular momentum vectors can vary with respect to the embedded galactic disks. The halo can consist of multiple kinematically distinct DM and baryonic components, e.g., streamers, subhalos and disk contributions, while overall being virialized \citep[e.g.,][]{diaz09}. 

Galactic disks can host single and double bars \citep[e.g., review by][]{shlo13}.
Barred galactic disks have been shown to lose their angular momenta to the host halos \citep[e.g.,][]{sell80,debatt00,atha03,marti06,bere07,villa09,villa10}, and do it mainly by resonance interactions \citep{atha03,marti06,dubi09}, as first derived by \citet{lynd72} and applied to barred disks by \citet{trem84}. However, these works have analyzed interactions with nonrotating host halos, clearly a tiny minority among the DM halo population. Alternatively, cosmological simulations which produce halos with various $\lambda$ still lack the necessary resolution to account for the resonant interactions. 

Recent modeling has indicated that the bar instability timescale shortens in disk galaxies with increased $\lambda$ \citep{saha13,long14}, confirming the theoretical prediction \citep{wein85}. Moreover, \citet{coll18a,coll18b} have investigated both dynamical and secular evolution of stellar bars in prograde\footnote{The {\it prograde} halo is used here in the sense of rotation in the direction of stellar disks and bars.} spinning halos, in the range of  $\lambda\sim 0-0.09$, and found substantial differences between the nonrotating and prograde spherical, oblate and prolate halos. 

In the present work, we have extended our analysis to the retrograde DM halos, i.e., when the halos counter-rotate with respect to the underlying barred stellar disks. We present a suite of models based on the same initial conditions, range of $\lambda$ up to $-0.09$, and perform a careful orbit analysis of these models to delineate the role of the resonances in the angular momentum redistribution.

Observations, theory, and numerical simulations have demonstrated that galactic disks reside in massive, responsive, DM halos. The tidal torque theory predicts that these halos gain most of their angular momentum by the action of gravitational torques during the phase of a maximal expansion \citep[e.g.,][]{doro70,white78a, fallefs}. It is less clear whether the halo's angular momentum is modified during the subsequent evolution, when virialized halos go through interactions, mergers and quiescent accretion processes \citep[e.g.,][]{porciani02}. For example, an increase in the angular momentum at this stage of evolution can be a transient phenomenon \citep[e.g.,][]{diaz07}.  

Galactic halos can include baryonic components in addition to the DM. The stellar halo motions can provide hints about the kinematics of the DM component. Baryons, in the form of gas and stars, can originate outside the halo, via accretion and mergers, or being injected by the embedded galaxies, either disks or ellipticals. In all cases, their origin will be imprinted on the kinematics and their angular momenta distributions. Stellar disks can possess cold counter-rotating components, which make up to 30\% of the disk mass \citep[e.g.,][]{kuij93,prada96,kann01,davi11,cors12,pizz14}, but they are outside the scope of this work, which focuses on the counter-rotation in the spheroidal DM component.

The halo can have distinct regions of angular momentum misalignment within itself.  \citet{diaz09} inspected the orientation of angular momentum vectors of different halo components and found that the it varies with time, e.g., their Figure\,19. The Milky Way has satellites in retrograde orbits \citep[e.g.,][]{lock03}, and numerical simulations show that the tidal streaming includes both material on prograde and retrograde orbits \citep[e.g.,][]{diaz10,paw11}. Note, \citet{dekel83} found that the angular momentum vectors of disks and halos can stay inclined to each other for a prolonged period of time and may be the cause of warps observed in galactic disks due to the tilted disk and halo.  

While statistics of angular momentum vectors orientation between galactic disks and their parent halos is not available at present, because the DM is not observable directly, some clues do exist. Numerical simulations have shown that counter-rotating disk components (with respect to the parent halo) can arise naturally in hierarchical clustering scenarios, even in the absence of merging \citep[e.g.,][]{algo14}. Moreover, as the observed disks can contain distinct components with anti-parallel angular momentum vectors --- one of these components should rotate against the DM halo. Our main question therefore is, how does the disk evolution change inside such a retrograde DM halo?

Early numerical simulations  have used rigid spherical or axisymmetric halos due to the lack of the computing power \citep[e.g.,][]{bour05}. Such halos are not able to absorb the angular momentum from the embedded stellar disk. Counter-rotating live halos or disks have been assumed not to be able to contribute to the angular momentum redistribution in the disk-halo systems and often compared to rigid halos \citep[][and refs. therein]{christ95}. However, this conclusion is based on assumption that the orbital structure of retrograde halos does not evolve. We find that certain aspects of this problem must be modified, and address this issue in the present work.

\citet{saha13} have simulated a counter-rotating halo of $\lambda = 0.05$ and found that such a halo delays the bar formation in the stellar disk. The stellar bar instability was slowed down compared to the nonspinning or prograde-spinning halos. However, these results are limited in scope in that they only analyzed model in the dynamical phase of bar evolution and only for one value of $\lambda$. The secular evolution of stellar bars in retrograde halos remains unknown.  

Does increasing the number of retrograde orbits in counter-rotating DM halos along the negative $\lambda$ sequence completely cut off all or almost all angular momentum transfer from the barred stellar disk to a the host halo? The near or complete absence of prograde orbits in the halo should limit the resonant coupling between DM and baryonic components. In this case, the small amount of material in the outer disk makes it more difficult for the inner disk (which is bar-unstable) to transfer large amount of angular momentum, perhaps limiting the disk expansion. We ask, if the retrograde halo orbits cannot resonate with the stellar disk, does this halo behave similarly to a rigid unresponsive halo and slows down the bar formation and growth?

This paper is organized as follows. Section\,2 presents the numerical aspects of our simulations, including initial conditions and orbital spectral analysis method. Section\,3 introduces our results on evolution of stellar bars in retrograde halos, and Section\,4 discusses our results. The last section summarizes our conclusions.

\section{Numerics}
\label{sec:Numerics}

We model stellar disks inside spherical \citet[hereafter NFW]{nava96} halos using the $N$-body part of the tree-particle-mesh Smoothed Particle Hydrodynamics (SPH/$N$-body) code GIZMO \citep{hop15}, which is a modified version of GADGET \citep{sprin05}. Our code units for mass, distance, and time are  $10^{10}\,M_\odot$, 1\,kpc, and 1\, Gyr. 

The DM halo contains $7.2\times 10^6$ particles and the stellar disk has $0.8\times 10^6$ particles. The halo mass is $M_{\rm h} = 6.3\times 10^{11}\,M_\odot$ and the disk mass is $M_{\rm d} = 6.3\times 10^{10}\,M_\odot$. Hence the ratio of DM to stellar particle mass ratio is near unity. For convergence analysis, we doubled the particle number in some models to create models with higher resolution which resulted in quantitatively similar evolution as those discussed here.

The opening angle, $\theta$, of the tree code and gravitational softening parameter are set to 0.4 and 25\,pc, respectively. Models presented here conserve energy at the level 0.05\% and angular momentum at 0.03\%, for the length of the 10\,Gyr runs.

\subsection{Initial Conditions}
\label{sec:ICs}

The initial conditions of the models follow the prescription of \citet{coll18a}, and are briefly restated here.

The halo density is given by the NFW profile,

\begin{equation}
\rho_{\rm h}(r) = \frac{\rho_{\rm s}\,e^{-(r/r_{\rm t})^2}}{[(r+r_{\rm c})/r_{\rm s}](1+r/r_{\rm s})^2},
\end{equation}
where $\rho(r)$ is the DM density in spherical coordinates, $\rho_{\rm s}$ is the fitting density parameter, and $r_{\rm s}=9$\,kpc is the characteristic radius, where the power law slope is $-2$, and $r_{\rm c}$ is a central density core where $r_{\rm c}=1.4$\,kpc. The Gaussian cutoff is applied at $r_{\rm t}=86$\,kpc for the halo.

The stellar disk is an exponential and we ignore the bulge potential. Its volume density is 

\begin{eqnarray}
\rho_{\rm d}(R,z) = \bigl(\frac{M_{\rm d}}{4\pi h^2 z_0}\bigr)\,{\rm exp}(-R/h) 
     \,{\rm sech}^2\bigl(\frac{z}{z_0}\bigr),
\end{eqnarray}
where $M_{\rm d}$ is the disk mass, $h=2.85$\,kpc is its radial scalelength, and $z_0=0.6$\,kpc is the scaleheight. $R$ and $z$ represent the cylindrical coordinates. The Gaussian cutoff is applied at $R_{\rm t}=6h\sim 17$\,kpc. Using these initial inputs, the halo-to-disk mass ratio within $R_{\rm t}$ is about 2. 
To initialize the halo velocities we freeze the disk potential and use a modified version of the iterative method from \citet{rodio06}, see also \citet{rodio09}. For a detailed description of technique applied see \citet{coll18a}. A short introduction to the iterative method follows.

We allow the halo to adjust to equilibrium velocities in the presence of of the frozen disk potential by allowing the DM particles to evolve from their initial positions and zero velocities for 0.3\,Gyr. Next, we use a nearest neighbor program to find the evolved particle that is closest to the position of an original particle at the start of the iteration. The original positioned particle is given the new velocity. We repeat the iterations until the halo velocities converge and obtain halo in virial and velocity equilibrium. The iteration routine required about 50 iterations to create a spherical NFW halo in equilibrium. To test the equilibrium, we ran the halos for an additional 3\,Gyr to verify that it is indeed in equilibrium.

The disk velocity profile depends on halo and disk mass distributions. We calculate the disk rotational
and dispersion velocities. The radial and vertical dispersion velocities assigned to the disk are

\begin{eqnarray}
\sigma_{\rm R}(R) = \sigma_{\rm R,0}(R){\rm exp}(-R/2h) 
\end{eqnarray}
\begin{eqnarray}
\sigma_{\rm z}(R) = \sigma_{\rm z,0}(R){\rm exp}(-R/2h), 
\end{eqnarray}
where $\sigma_{\rm R,0} = 120\,{\rm km\,s^{-1}}$ and $\sigma_{\rm z,0} = 100\,{\rm km\,s^{-1}}$. 
The Toomre's parameter was calculated to have a minimum of $Q\sim 1.6$ at $R\sim 2.4 h$. As expected, $Q$ increases towards the center and the outer disk.

The above procedure creates a halo with cosmological spin parameter $\lambda =0$. To spin up the halo in the retrograde direction, we have reversed the tangential velocities of a fraction, $f$, of prograde particles (with respect to the rotation of the disk). $f$ is increased to create halos of increasingly negative $\lambda$. Here we present halos with a range $\lambda\sim 0-0.09$. The new velocity distributions maintains the solution to the Boltzmann equation and do not alter the velocity profile \citep{lynd60,wein85,long14,coll18a}.  For axisymmetric systems, the invariance with velocity reversals is a direct consequence of the \citet{jeans19} theorem \citep[see also][]{binn08}.

Therefore, we have produced a suite of disk-halo models which differ only in the spin, $\lambda$. Following the notation of \citet{coll18a} and \citet{coll18b} the models are labeled as $P$ if they are prograde and $R$ if they are retrograde, and then multiplied by 1,000. For example,  the standard model, P00,  is the nonspinning halo with $\lambda = 0$, and R60 is the halo with retrograde rotation and $\lambda = 0.06$.

\subsection{Orbital Spectral Analysis Method}
\label{sec:spectral}

We analyze the orbital structure of our disk-halo models and examine the role of resonant angular momentum transfer by using the orbital spectral analysis method \citep{binneyspergel82,atha03,marti06,dubi09,coll18b}. Using Fourier analysis, we determine the angular velocity, $\Omega$, and the radial epicyclic frequency, $\kappa$, for stellar and DM orbits which resonate with the stellar bar pattern speed, $\Omega_\textrm{b}$. This is performed in the frozen total potential. In the bar frame, we construct the dimensionless frequency $\nu\equiv (\Omega-\Omega_\textrm{b})/\kappa$ and plot the distribution of orbits with $\nu$. Each stellar or DM orbit has been evolved for $30-50$ orbits. For more information see \citet{coll18b}.

\section{Results}
\label{sec:results}

We present our results dealing separately with the stellar and DM components in the counter-rotating DM halos. We start with tracking the stellar bar evolution and follow up with the DM response in these models.

\begin{figure*}
\centerline{
\includegraphics[width=\textwidth,angle=0] {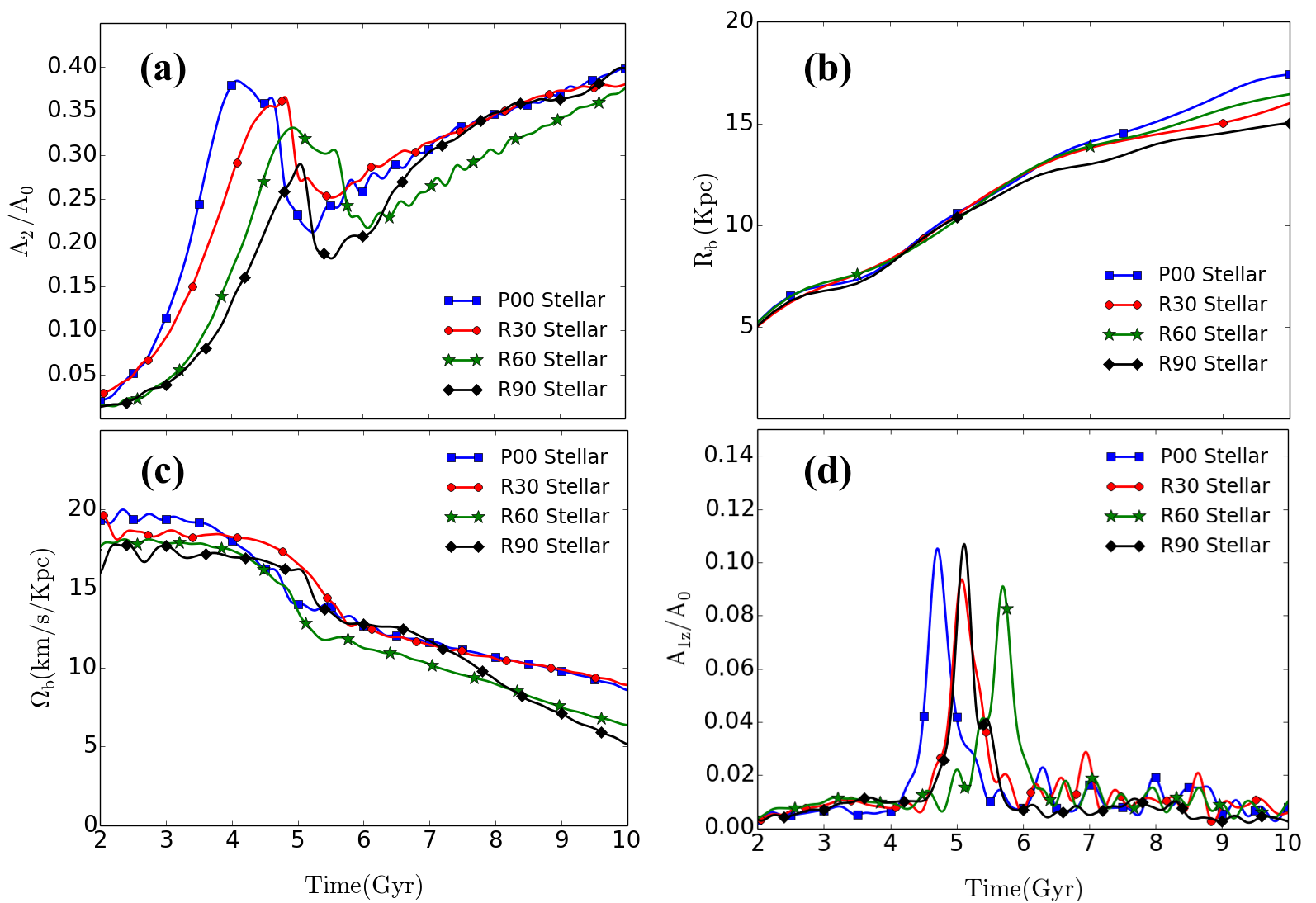}}
\caption{(a) Fourier amplitude, $A_2$, normalized by the monopole term, $A_0$, evolution of the stellar bars along the retrograde $\lambda$ sequence in our models; (b) Bar length, $R_\textrm{b}$; (c) Evolution of the bar pattern speeds, $\Omega_\textrm{b}$, in retrograde halos; (d) Vertical buckling amplitude, $A_\textrm{1z}$, of stellar bars, normalized by $A_0$. The fiducial model, P00, represented by a blue line in all plots, has been added for comparison.}
\label{fig:stellar}
\end{figure*}

\subsection{Retrograde Stellar Models Evolution}
\label{sec:stellar}

We measure the stellar bar strength amplitude, $A_{2}$, the bar length, $R_\textrm{b}$, the pattern speed, $\Omega_\textrm{b}$, and finally the vertical buckling amplitude, $A_\textrm{1z}$. Figure\,\ref{fig:stellar}a exhibits the evolution of the stellar bar amplitudes in retrograde models. The P00 model evolution has been added for comparison. A number of conclusions can be drawn by comparing this Figure with behavior of $A_2$ in Figure\,1 of \citet{coll18a}. First,  arranging  the prograde and retrograde models from largest positive $\lambda$ to the most negative one, the bar instability timescale increases monotonically. 

Next, while stellar bars in prograde models display approximately the same maximal pre-buckling amplitude in the pre-buckling stage, as Figure\,\ref{fig:stellar}a shows, there is a gradual decrease in this amplitude for retrograde models.

Third, all stellar bars buckle and reduce their amplitude abruptly, both in the prograde or retrograde models. The prograde models display progressively lower minimum in $A_2$ with $\lambda$, and this trend continues in the retrograde models. But the drop in $A_2$, i.e., $\Delta A_2$, is much less dramatic in the retrograde models. 

Fourth, in the secular stage of evolution, the amplitudes of bars in retrograde models experience a monotonic growth, with rather minor differences in the bar strength. Contrary to this, the prograde models show a much more complex behavior, which includes essentially the bar dissolution for larger prograde (i.e., positive) $\lambda$ \citep[][]{coll18a,coll18b}. This constitutes probably the largest and most profound difference in the evolution of the prograde and retrograde models. During the secular phase, the stellar bars experience a healthy growth in the retrograde models, but those in the prograde models do not grow after the buckling, and appear nearly dissolved.

The stellar bar sizes have been determined from extension of the major axes of the $x_1$ orbits. These orbits are the main orbits which support the bar and are elongated along the bar. They populate the region between the corotation and the Inner Lindblad resonance (ILR). Because finding these orbits can be time consuming, we have confirmed the bar length using an alternative method, by measuring ellipticity profiles of their isodensity contours in the $xy$-plane. The bar has been assumed to extend to the point where ellipticity has decreased by 15\% from its maximum \citep[e.g.,][]{marti06,coll18a}. We plot evolution of the stellar bar length in Figure\,\ref{fig:stellar}b. 

The continued stellar bar growth is indicative of its braking against the halo and transferring its angular momentum to the halo and outer disk. Each stellar bar in retrograde halos grows in size for the entire run. During the secular stage, we observe an increasing difference between the bar sizes. The R90 stellar bar grows more slowly than the P00 stellar bar, and at $t=10$\,Gyr, this bar is about 20\% shorter than the P00 bar. 

Evolution of the bar pattern speeds in all retrograde models and in P00 is given in Figure\,\ref{fig:stellar}c. We observe that angular momentum transport, which is facilitated by the bar braking against the outer disk and the halo, is not inhibited by an increase in the retrograde halo spin. In each model, the stellar bar slows down over the entire simulation. We can compare this evolution to that of the stellar bars in the prograde models where increasing $\lambda$ leads to a decrease in the bar amplitude and the near dissolution of the
bar, leading to a much weaker bar braking and slowdown.

We plot the Fourier amplitude of the vertical buckling, $A_\textrm{1z}$, in the $rz$-plane in Figure\,\ref{fig:stellar}d. The  $A_\textrm{1z}$ is normalized by the monopole term, $A_0$. The maximum amplitude of the buckling instability does not depend on the $\lambda$, but the time of buckling does depend, as is also evident from evolution of $A_2$.

\begin{figure}
\centerline{
\includegraphics[width=0.5\textwidth,angle=0] {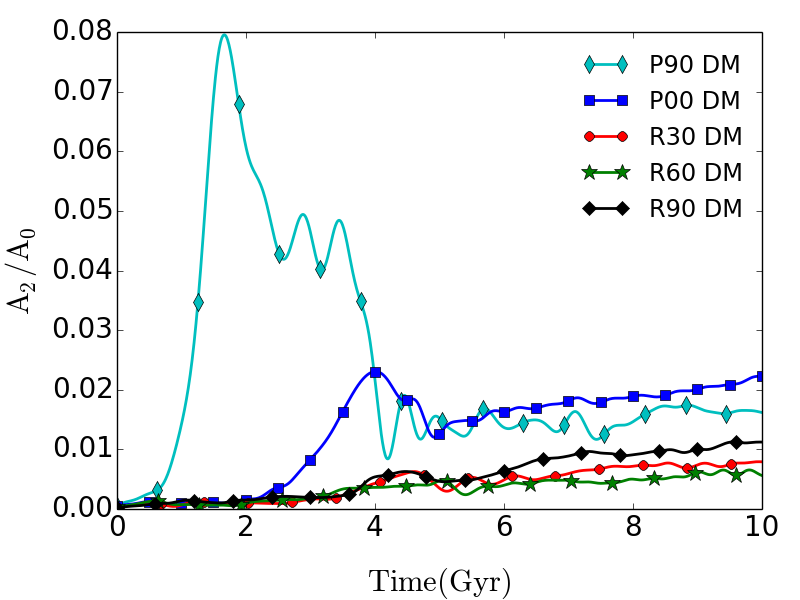}}
\caption{Fourier amplitudes, $A_2$, of the DM response to the stellar bars along the negative $\lambda$ sequence in the retrograde models. The amplitude is normalized by the monopole term, $A_0$. Models P00 with $\lambda=0$ and P90 with $\lambda=0.09$ have been added from \citet{coll18a,coll18b} to provide the overall picture of the DM response in prograde and retrograde halos. The same integration limits in $r$ and $z$ have been used as for the stellar bars for an accurate comparison. Note the change in scale of the $y$-axis when compared to the stellar bars in Figure\,\ref{fig:stellar}a.}
\label{fig:dma2}
\end{figure}

\begin{figure}
\center
\centerline{
\includegraphics[width=0.48\textwidth,angle=0] {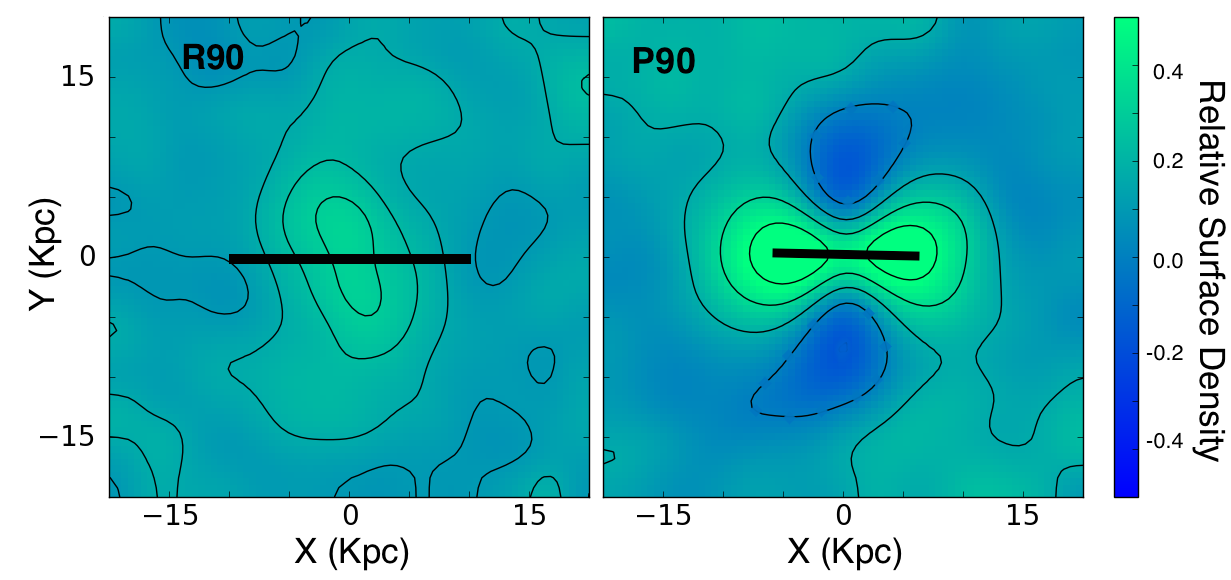}}
\caption{Projected density response of DM halos in retrograde (R90) versus prograde (P90) DM halos. The response is displayed in the pre-buckling phase, when the stellar bars are close to their maximal strength, i.e., at $t\sim 5$\,Gyr (R90) and $\sim 1.8$\,Gyr (P90). Shown is the ratio of the projected DM density on the $xy$-plane within $|z| < 3$\,kpc over the same at $t= 0$, and subtracting unity from the ratio. The contour levels are given in the color palette. Positions of stellar bars are delineated by the straight horizontal line. The P90 model is from \citet{coll18b}. The positive contours are black solid lines and the negative ones are dashed lines. The outline of density enhancements and deficiencies delineate the DM response, including the DM gravitational wakes. Note that both extension and projected density perturbation amplitude of the DM response vary with $\lambda$.
}
\label{fig:DMrespo}
\end{figure}

\subsection{Dark Matter Response in Retrograde Models:
Orbital Reversals}
\label{sec:reversals}

The DM response to the stellar bar perturbation is shown in Figure\,\ref{fig:dma2}. For a comparison, two models have been added --- the $\lambda=0$ model, P00, and the $\lambda=0.09$ prograde model, P90.  Note that, as usually, the Fourier amplitude of $m=2$ mode in the DM  is much weaker than the stellar amplitude. Avoiding the semantic discussion pertaining to what $A_2$ value defines a `DM bar',  we refer to the DM response in all models as a DM bar. We do note that the DM response in retrograde halos is distinctively weaker than its response in the prograde models. It tumbles in the direction of the stellar bar. In other words, the DM response follows the stellar bar, thus propagating against the spin of the DM halo. Furthermore, as we show below, the DM response lags the stellar bar by almost $90^o$
degrees.

During the dynamical phase of the bar instability, we observe a substantial difference in the DM response. The timescale of the bar instability becomes more prolonged with decreasing $\lambda$ from 0.09 to -0.09. Hence, in pre-buckling evolution of prograde and retrograde models, we observe a clear hierarchy, from P90 to to P00, and to R30, followed by R60 and R90. But what is most interesting is the behavior of the amplitude, $A_2$, of the DM response, whose pre-buckling maximum decreases from 0.08 for P90 to 0.024 for P00 and to 0.008 for R90 --- overall by a factor of 10. In comparison, the {\it stellar} bar pre-buckling amplitudes stay about the same for $\lambda = 0-0.09$, and display a small decrease for the retrograde models, down to $\lambda= -0.09$, as shown by Figure\,1 of \citet{coll18a}. 

During the secular phase of evolution, the DM response in retrograde halos appears much weaker than in the prograde models. An important point to be emphasized now and to which we shall return in the subsequent sections, is the break in the monotonic evolution of $A_2$ DM amplitudes in the secular phase. This is especially noticeable in the amplitude for R90 model. It has been naively expected to be the weakest in Figure\,\ref{fig:dma2}, based on the evolution of other retrograde models, yet it evolved and became the strongest instead. The explanation for this phenomenon is not a trivial one, and will be addressed below. 

\begin{figure*}
\centerline{
\includegraphics[width=\textwidth,angle=0] {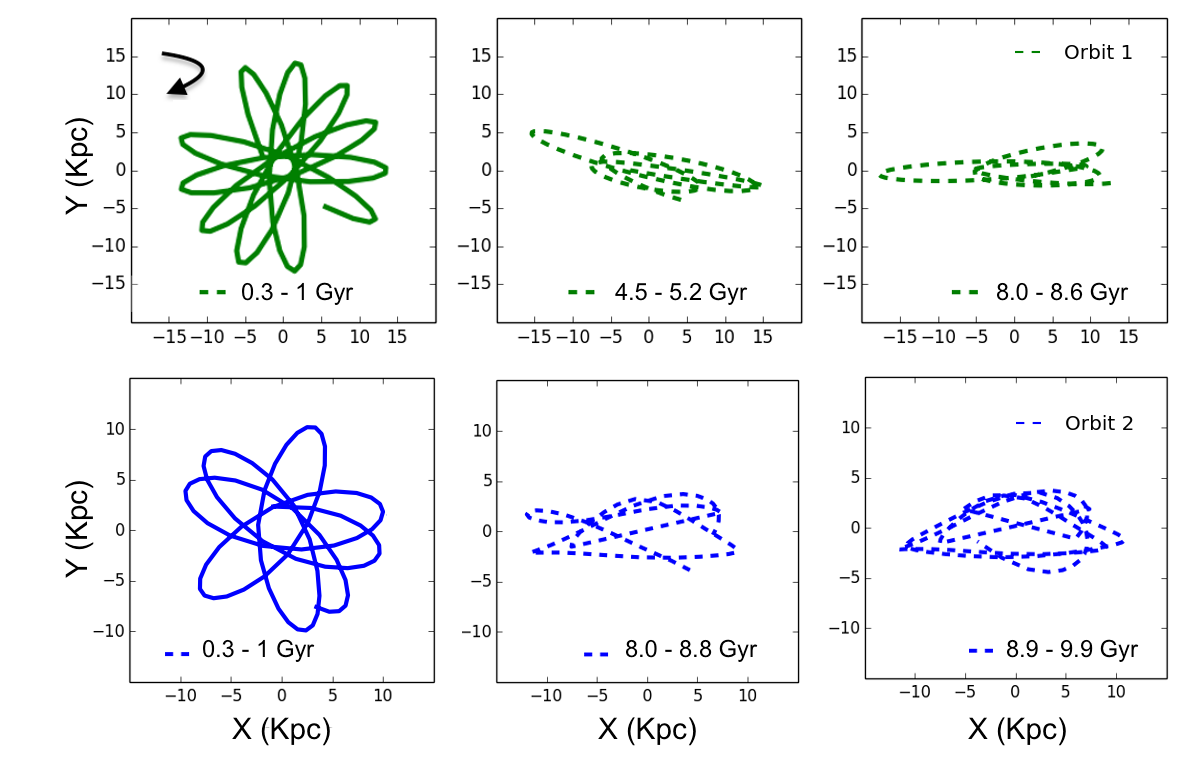}}
\caption{The randomly chosen two DM orbits of R90 model, initially counter-rotating with the disk (solid lines) and finally corotating (dashed lines) with it. Both orbits have been evolved for three revolutions in the live potential in each of the three frames (time periods are shown). Before the stellar bar appears (left frames), during the strong bar before its buckling (upper middle frame) and when the stellar bar strengthens during secular evolution (lower middle frame). Orbits remain captured towards the end of the run, during secular evolution (right frames). The middle and right frames are shown in the stellar bar frame of reference, i.e., with the bar pattern speed subtracted. The stellar bar is positioned horizontally. The upper left frame curled arrow shows the direction of precession of both orbits before the bar instability. This direction is reverse when the orbits are captured. The orbits remain trapped by the stellar bar in the last two frames.  
}
\label{fig:reversal1}
\end{figure*}
\begin{figure}
\centerline{
\includegraphics[width=0.5\textwidth,angle=0] {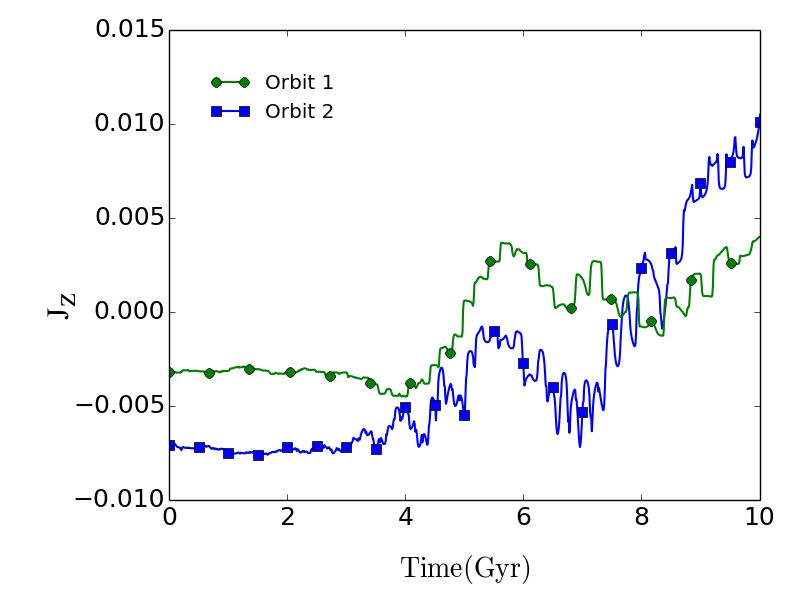}}
\caption{Evolution of the angular momentum, $J_\textrm{z}$, of two DM orbits, projected on the $xy$-plane, which are mapped in Figure\,\ref{fig:reversal1}. The colors of individual orbits in these figures are matched. The orbits start as retrograde, i.e., having negative angular momentum with respect to the stellar disk and bar rotation. Both orbits switched their streeming and precession directions when trapped by the stellar bar, either before buckling or during the secular phase of evolution. Both orbits are briefly released back to the disk and captured again. The units of angular momentum on the $y$-axis are $M_\odot$\,$\textrm{km}\,\textrm{s}^{-1}\,\textrm{kpc}^{-1}$.
}
\label{fig:reversalJ}
\end{figure}

To emphasize the difference between the DM response in prograde and retrograde models, we display the projected DM density onto the $xy$-plane in Figure\,\ref{fig:DMrespo}. Note the dramatic change in the morphology of this response between $\lambda=-0.09$, 0, and $+0.09$. The prograde model displays a response aligned with the stellar bar. While the retrograde model shows response which is nearly $90^o$ trailing the stellar bar. This difference underscores the importance of the CR resonance in the prograde models versus the ILR in the retrograde models. We discuss the importance of these resonances in this and the following sections.  

A more careful study of Figure\,\ref{fig:dma2} reveals a more complex behavior of DM bar in the R90 model compared to other retrograde models. It displays a
faster growth of $A_2$ after $t\sim 6$\,Gyr and associated stronger braking at the same time. The DM bar in the R60 model shows a weaker version of this evolution, when its DM bar strength increases to match the rival R30. Analyzing this behavior, we came across an unexpected process, which was not discussed in the literature so far --- this process sheds a new light on stellar bar evolution and DM response in retrograde models. It involves angular momentum exchange between the stellar bar and the retrograde DM orbits. As a result of this interaction, the DM orbit gains angular momentum and reverses its direction of streaming and precession.  We term this process as {\it orbit reversal}. To understand the dissentic evolution of the R90 model, and to a lesser degree of all the retrograde models, we correlate it with the DM orbit reversal. During this process the DM orbit and the halo as a whole absorb angular momentum from the disk. We first demonstrate that the orbit reversals do occur.

Figure\,\ref{fig:reversal1} displays two examples of such reversals in the R90 model. We pick randomly two orbits that started as retrogrades at $t=0$ and finished as progrades at $t=10$\,Gyr. To limit the search volume, we only look at the orbits within $R < 20$\,kpc and $|z| < 10$\,kpc, i.e., close enough to the disk. 

For each orbit, we start by choosing three characteristic times. The first one is close to the starting time of the run, at $t = 0.3$\,Gyr, when the disk is axisymmetric, and integrate the orbit until 1\,Gyr. The next time period for the first orbit is picked when the stellar bar is close to its maximal strength before the buckling, at $t = 4.5$\,Gyr, and integrate it till $t\sim 5.2$\,Gyr. Lastly, we choose the time close to the end of the run, during the secular phase, at $t = 8$\,Gyr and integrate it to $t\sim 8.6$\,Gyr. At each time period, we follow this orbit for about three revolutions in the {\it live} potential of the system. The second example displays a reversal DM orbit at the time periods of $t\sim 0.3-1$\,Gyr, $8-8.8$\,Gyr, and $8.9-9.9$\,Gyr. 

The left frames of Figure\,\ref{fig:reversal1} display the orbital precession, the rosette, in the rest frame, for both orbits. These rosettes have strfeaming and precession in the same direction. Their angular momentum, $J_{\textrm z}$, being negative, are shown on Figure\,\ref{fig:reversalJ} with the associated colors in Figure\,\ref{fig:reversal1}. The middle frames, which are in the reference frame of the stellar bar (being horizontal), show the same orbits being trapped by the bar and already reversed their direction of streaming and precession. Their $J_\textrm{z}$ is positive at this time. The right frames show the orbits being trapped by the bar towards the end of the run. 

Figure\,\ref{fig:reversalJ} reveals that the first orbit was captured early and briefly released by the bar into the disk after buckling and captured again. The second orbit was not captured by the stellar bar until late in the evolution. But it interacted strongly with the bar before hand, as its $J_\textrm{z}$ oscillated widely. Acquiring positive angular momentum does not assure that the orbit will remain prograde.

\begin{figure}
\centerline{
\includegraphics[width=0.5\textwidth,angle=0] {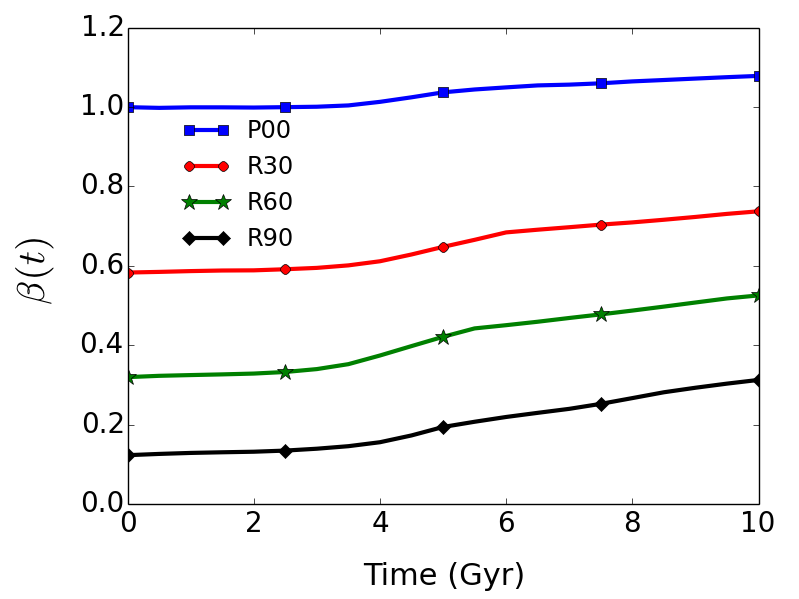}}
\caption{The ratio of prograde to retrograde DM orbits, $\beta(t)$, within $R < 20$\,kpc and $|z| < 10$\,kpc for 
retrograde models and for P00 in comparison. While the fraction of prograde orbits in P00 increases little during the
evolution, more negative $\lambda$ leads to the increase in $\beta$ with time, due to the retrograde DM orbit
reversals (see the text for more information).}
\label{fig:ratio2}
\end{figure}

The frequency and importance of these DM orbital reversals can be quantified. For this purpose, we calculated the fraction of retrograde DM orbits as a function of time for each of the retrograde models, and for the P00 model for comparison. All DM orbits within the region of $R < 20$\,kpc and $|z| < 10$\,kpc have been counted.

\begin{figure*}
\centerline{
\includegraphics[width=\textwidth,angle=0] {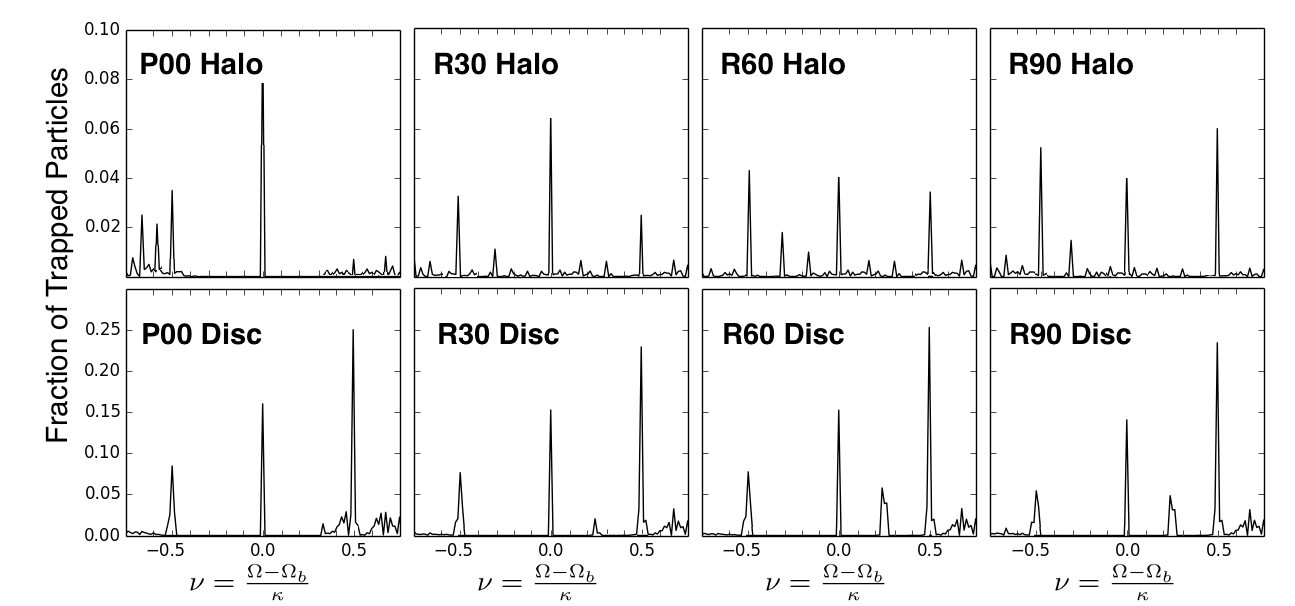}}
\caption{The orbital spectral analysis of disk and halo orbital structure: the negative $\lambda$ sequence of resonance trapping for DM halos (top) and stellar disks (bottom). The $x$-axis gives the normalized frequency, $\nu\equiv (\Omega-\Omega_{b})/\kappa$ (see definition in the text). The $y$-axis is the fraction of orbits trapped at each of the major resonances, the ILR ( $\nu=0.5$), the CR ($\nu=0.0$), and the OLR ($\nu=-0.5$). The frequencies are binned in $\Delta\nu=0.01$. The chosen time for spectral orbital analysis is at $t=8$\,Gyr for all models.}
\label{fig:spectra}
\end{figure*}

Figure\,\ref{fig:ratio2} presents evolution of the ratio of prograde-toretrograde orbits, $\beta(t)$, in the inner halo of $R < 20$\,kpc and $|z| < 10$\,kpc. This region contains DM orbits that can possibly interact and be trapped by the stellar bar. All the curves in this Figure are relatively flat before the stellar bar acquires it strength, indicating that the orbit reversals in the DM halo is due to the stellar bar and not due to the instability in the DM halo. In fact we have previously run diskless spinning halos and found them to be completely stable and their density and velocity distributions show no evolution \citep{coll18b}. 

During the buckling and the subsequent secular evolution of stellar bars in retrograde halos, we observe that
$\beta(t)$ increases with time (Figure\,\ref{fig:ratio2}). This increase is more substantial with $\lambda$
becoming more negative. For R90 model, the initial $\beta$ is 0.12, and it increases to 0.31 after the buckling.
The comparison P00 model displays a minimal change --- from 1.0 to 1.08 only. The sequence of retrograde models
shows a monotonic increase of orbital reversals, from P00 to R90. In fact, $\beta$ at 10\,Gyr in R90 is equal to
the initial $\beta$ of these orbits in the R60 model. Hence, the difference between these models has been erased over the evolution time, as seen in Figure\,\ref{fig:ratio2}. The greatest rate of increase in the DM prograde orbits is observed in R90 after $t\sim 5$\,Gyr. In other words, exactly when the stellar bar in R90 increases in strength. It seems plausible, that this effect of orbital reversals is responsible for the increased strength of stellar bar, as additional orbits in the DM halo become resonant with the stellar bar.

In the subsequent analysis in section\,\ref{sec:orbital_analysis}, we inquire whether the reversed DM orbits are indeed trapped by the stellar bar. A substantially increased fraction of prograde DM particles would be able to resonate with the stellar bar, amplify the angular momentum transfer rate from the disk to the halo, and, as a result, strengthen the stellar and DM bar components as seen in Figure\,\ref{fig:dma2}. 

\subsection{Orbital Spectral Analysis During the Secular Evolution}
\label{sec:orbital_analysis}

For a more detailed look at the behavior of retrograde models, we perform the orbital spectral analysis to determine the fraction of orbits trapped at each resonance, and to clarify the role of the orbital reversals in this process along the negative $\lambda$ sequence.
The chosen time for this analysis is identical for all models, at $t=8$\,Gyr. This time is long after buckling, i.e., during the secular growth phase of stellar bars. For comparison, \citet{coll18b} performed this analysis for the prograde models before buckling. The reason for this was that for prograde models, increasing $\lambda$ essentially destroys the stellar bars for $\lambda\gtorder 0.06$. Contrary to this, the retrograde models presented here, experience a strong growth of stellar bars after the buckling.

The result of orbital spectral analysis for each model is shown in Figure\,\ref{fig:spectra}. We plot the fraction of trapped orbits on the $y$-axis versus dimensionless frequency, $\nu\equiv (\Omega-\Omega_\textrm{b})/\kappa$, on the $x$-axis. The peaks correspond to frequencies where stellar and/or DM particles are trapped by the resonances. The bottom frames show the stellar disks and the top frames show the DM halos with increasingly negative $\lambda$ sequence. In all models, the highest spikes correspond to the familiar resonances: the inner Lindblad resonance, ILR at $\nu=0.5$, the corotation resonance, CR at $\nu=0$, and the outer Lindblad resonance, OLR at $\nu=-0.5$. The stellar bar strength at this time is approximately the same in all retrograde models. 

\begin{figure*}
\centerline{
\includegraphics[width=\textwidth,angle=0] {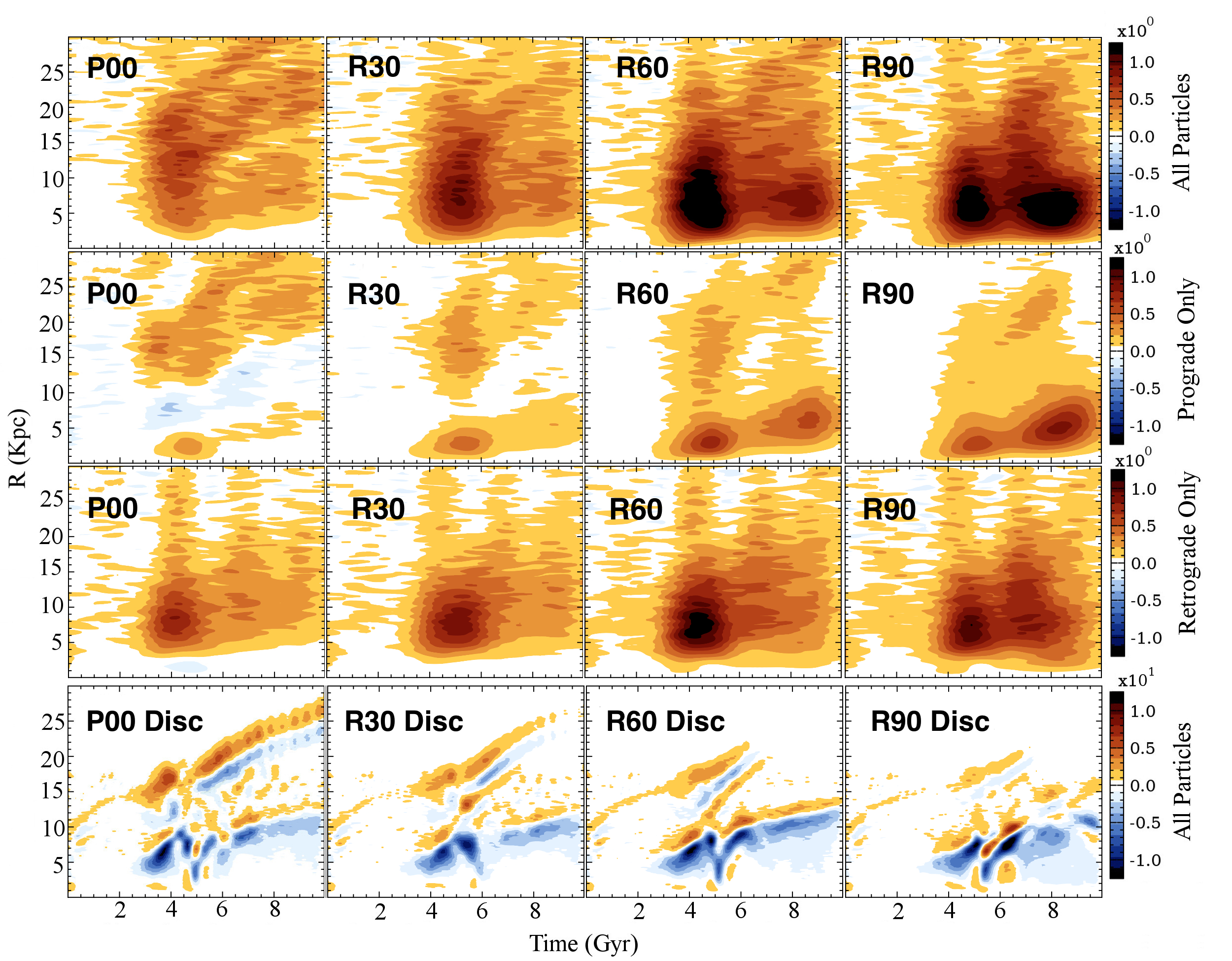}}
\caption{Rate of the angular momentum transfer, $\dot J$ in the retrograde disk-halo systems. The top three rows show $\dot J$ emission and absorption by prograde and/or retrograde DM halo orbits, as a function of a cylindrical radius $R$ and time, along the retrograde $\lambda$ sequence. The color palette corresponds to gain/loss rates in $J$, i.e., red/blue,  using a logarithmic scale in color. The cylindrical shells are binned at $\Delta R = 1$\,kpc and extend to $z=\pm 10$\,kpc. The top row includes both prograde and retrograde orbits in the DM halo. The second row --- only the prograde orbits, and the following row --- only the retrograde orbits. The bottom row shows prograde and retrograde orbits in the disk. The unit of angular momentum transfer rate used in the color palette is $M_\odot$\,$\textrm{km}\,\textrm{s}^{-1}\,\textrm{kpc}^{-1}$.}
\label{fig:jdot}
\end{figure*}

The bottom row of Figure\,\ref{fig:spectra} shows that the efficiency disk orbit trapping at the ILR does not change with $\lambda$. This resonance resides deep inside the stellar bar and is the dominant resonance which 'emits' the angular momentum by the disk. On the other hand, as we move along the $\lambda$ sequence, a monotonic decline in the trapping ability of the OLR can be observed. For example, for R90, the OLR appears to trap about half of the orbits compared to the OLR in P00 model. There is also a smaller reduction in the trapping efficiency of the CR, with increasingly negative $\lambda$. In addition, a peak develops between the CR and ILR of the disk, at $\nu \sim 0.25$ --- the Ultra-Harmonic 1:4 resonance (UHR). This result implies that increasingly retrograde $\lambda$, gradually weakens the trapping ability of the outer disk resonances.  

The top row of Figure\,\ref{fig:spectra} displays an opposite trend in the DM halos to that observed in the stellar disks. In the P00 halo, we see that the OLR is weak and the ILR is very weak. The inner resonance, the ILR increases the efficiency of the DM orbit trapping, while the CR shows a decrease, with increasingly negative $\lambda$. The OLR exhibits a moderate increase in its trapping ability. In the R60 and R90 models, for example, the ILR and OLR rival the trapping efficiency of the CR. The bottom row shows unchanged activity in the ILR and the CR, and decreased trapping efficiency by the OLR.

In a way, the stellar bar displays the trend of preferentially trapping the DM halo orbits rather than stellar orbits in the outer disk. The P00 model shows a strong DM peak at the CR, which is thought to be the most important resonance for absorbing the angular momentum by the halo. With increasingly negative $\lambda$, this resonance becomes less important compared to the OLR and ILR. We find that this decrease in the CR trapping is associated with the increase in the retrograde particle fraction with $\lambda$ in the DM halo.

Though the importance of each resonance in the halo changes with $\lambda$, we note that the total fraction of trapped orbits remains similar, $\sim 20\%$ within the sampled region. This trapping process distinguishes retrograde halos from the rigid unresponsive halos and shows that this system is more complex than previously thought. Rigid halos do not respond to the torques of stellar bars at all, while we find that the disk is adept at trapping orbits even in retrograde halos. For prograde halos this was pointed out by \citet{atha02}, using DM halos which were initially 
non-rotating.

This varying efficiency of resonance trapping for the disk and halo orbits with the retrograde $\lambda$ sequence, demonstrates that the details of angular momentum transfer in these models must vary as well. We take a closer look at the angular momentum transfer between disks and halos in the following section.

\subsection{Rates of Angular Momentum Transfer}
\label{sec:rate_of_J}

An alternative way of analyzing  the interactions between retrograde halos and embedded disks, without referring to the resonances, is to visualize the flow of angular momentum in a galaxy using the method prescribed in \citet{villa09}, and implemented elsewhere
\citep[e.g.,][]{long14,coll18a, coll18b}. The halo and disk are binned into cylindrical shells of $\Delta R = 1$\,kpc. We create two-dimensional maps of the rate of change of $J$ in each shell as a function of $R$ and time. These ${\dot J}$ maps are then assigned a color palette, where a gain of angular momentum is given in red and a loss of angular momentum in blue.  The color palette has been normalized separately for the disk and the halo.

In the top row of Figure \ref{fig:jdot},  $\dot J$ --- the rate of the angular momentum transfer has been calculated from and to the DM halo, for models along the retrograde $\lambda$ sequence. The P00 halo displays a nearly pure absorption of $J$ by the halo. The weak
emission is related to the Ultra-Harmonic resonance (UHR). Three resonances are clearly seen in this frame --- the ILR, CR, and OLR, which also appear in the $\dot J$ map of the P00 disk. In the P00 model, the highest $J$ transfer happens close to the time of buckling, where the stellar bar is the strongest, and where we see the deepest emission and absorption.

Increasing the retrograde $\lambda$, along the top row in Figure\,\ref{fig:jdot}, a stark contrast appears between the halo models. In the inner $R < 10$\,kpc of models with larger (negative) $\lambda$, two deep absorption features are seen. The first one corresponds to the buckling of the stellar bar associated with high $\dot J$ transfer to the halo. The second deep absorption feature appears in R60 and increases in strength to R90. Note that this feature appears after $t\sim 6$\,Gyr, and so can be associated with the reversal of DM orbits discussed earlier. The version of this figure for the prograde spinning halos is shown in \cite{coll18b}. We note that increasing $\lambda$ in the retrograde direction shows that the DM halos only absorb the angular momentum. The emission features are absent in the top row of Figure\,\ref{fig:jdot}. But these emission regions can be seen prominently in halos rotating in the prograde direction, e.g., Figure\,10 of \cite{coll18b}.

The second and third rows of Figure \ref{fig:jdot}, display the rate of transfer of angular momentum for the prograde and retrograde DM orbits separately.  The P00 halo has 50\% of orbits rotating with the disk, and these prograde orbits gain and lose $J$, clearly following the resonances produced by the stellar bar. In the retrograde orbit plot for the P00 model, only absorption can be seen. For different models with increasing (negative) $\lambda$, the fraction of prograde orbits decreases, and only absorption is visible for the prograde orbits. The ILR resonance appears more important to prograde orbits in these halos, and the gain of angular momentum increases with retrograde $\lambda$. 

The double peaks in $\dot J$ are clearly visible in the prograde DM $\dot J$ maps, and the second peak appears after $t\sim 6$\,Gyr.

The third row of Figure \ref{fig:jdot} displays only the retrograde orbits, where a gain of angular momentum is
pronounced in all models. Notably, this gain in angular momentum increases with $\lambda$, which corresponds to
the increase in prograde orbit fraction, as seen in Figure\,\ref{fig:ratio2}. The second maximum in $\dot J$ is weaker here and happens slightly early in time than for the prograde orbits. 

The final row of Figure\,\ref{fig:jdot}, displays the rate of angular momentum transfer for all orbits found in the stellar disk. In the P00 disk, the importance of the resonances can be clearly observed, and the $J$ transfer reaches larger $R$, as the bar grows in size and slows down. As we shall see later, the size of the disk correlates with the $J$ transfer as well. The stellar bars differ in length by not more than 20\% (Figure\,\ref{fig:stellar}b), So they do not differ dramatically from each other. But the $J$ transfer appears quite different when mapped using this method. 

The OLR in the P00 disk extends to $25$\,Kpc, as seen in the color map of Figure\,\ref{fig:jdot}.  For comparison, along the retrograde $\lambda$ sequence, the OLR is stunted and stops well before 20\,kpc in R60 and R90 models. 

Stellar disks in prograde models are shown in the  Figure\,6 of \citet{coll18a} and analyzed in \citet{coll18a,coll18b}.

The second deep absorption feature in the DM halos, that gets stronger with increasing $\lambda$, coincides with the emission feature within central $R <5 $\,kpc in the disk, which gets stronger with $\lambda$. We shall return to this issue in the following section.

\begin{figure*}
\centerline{
\includegraphics[width=\textwidth,angle=0] {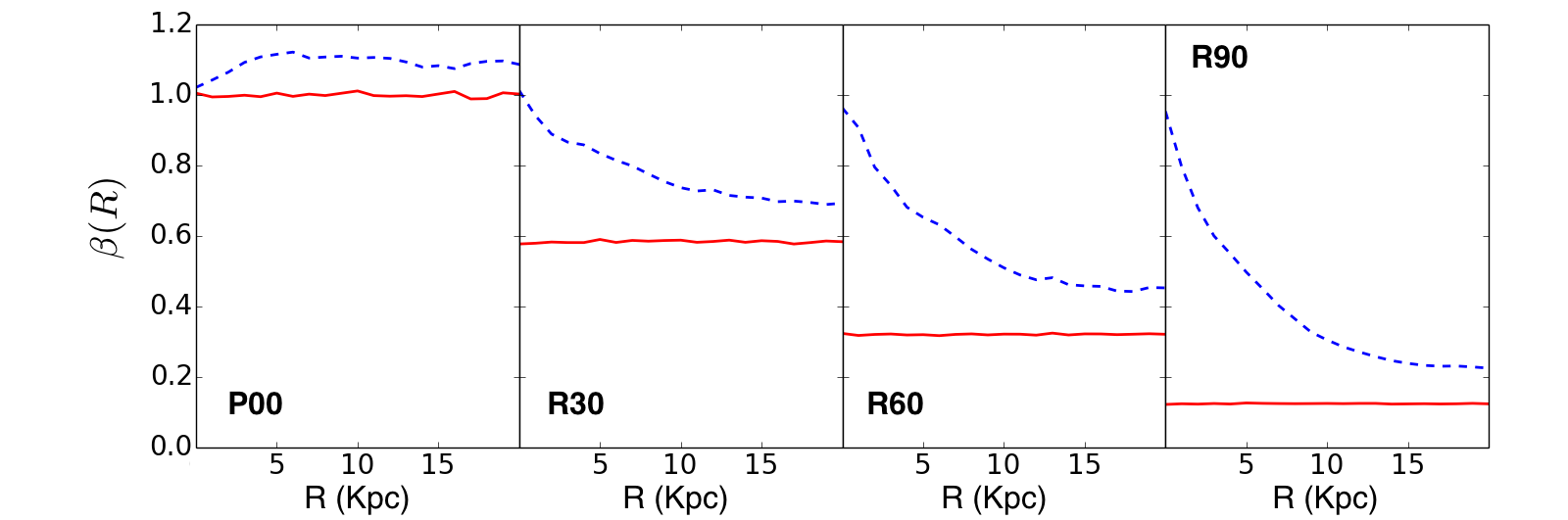}}
\caption{Ratio of prograde to retrograde orbits, $\beta(R)$, as a function of $R$, for the inner halo, shown here for $R < 20$\,kpc, measured at $t = 0$ (solid line) and $t = 10$\,Gyr (dotted line), for different models with increasing retrograde $\lambda$.}
\label{fig:ratio}
\end{figure*}

\section{Discussion}
\label{sec:discussion}

We have analyzed the evolution of stellar disks embedded in counter-rotating DM halos over time period of 10\,Gyr. We focused on the dynamical and secular evolution of stellar bars in these systems, on the DM response to the stellar bar perturbation, and on the flow of angular momentum between the halos and the embedded disks. The range of the counter-rotating DM halo spin used is $\lambda = 0-0.09$. Finally, we have compared the evolution in the prograde and retrograde halos. After summarizing our results, we discuss their corollaries and additional questions they bring.

Our main results are as follows.  First, we find that the maximum strength and size of stellar bars is only moderately affected by the retrograde halo spin during dynamical and secular phases of evolution. This is in a stark contrast with the prograde sequence \citep{coll18a,coll18b}. The largest difference comes from larger $\lambda$ --- prograde or retrograde. While the prograde models are characterize with dissolution of stellar bars in the secular stage of the evolution, no such trend is found for the retrograde models --- all bars here show a healthy growth until the end of the simulations.

Second, the stellar bar amplitudes during the bar instability in the retrograde halos form an extension to the sequence of bar evolution in prograde halos by delaying the bar instability. However, the amplitude peaks of the retrograde models appear more crowded in time --- the delay in prolonging the bar instability saturates.   

Third, we have performed the orbital spectral analysis on retrograde models in order to quantify the overall trapping efficiency of DM orbits by the stellar bar. We find that trapping is not affected along the counter-rotating $\lambda$ sequence, i.e., stellar bars trap $\sim 20\%$ of the DM halo particles in the sampled region of the inner halos, despite the increasing fraction of retrograde orbits in the initial conditions.  Again, this is contrary to the prograde $\lambda$ sequence of disk-halo models which exhibits a strong effect of the halo spin on stellar bar evolution.

Fourth, although the overall trapping ability of the DM orbits is not affected along the retrograde $\lambda$ sequence, the trapping by the individual resonances does vary. For example, the ILR and OLR resonances in the DM halos become progressively more important with increasing $\lambda$, while the stellar disk shows a decrease in trapping ability of stellar orbits by the OLR resonance. The CR and the ILR resonances do not change. The UHR appears in the disk and becomes stronger with $\lambda$. We also have measured the angular momentum, $J$, absorbed or emitted by each of the resonances during the secular evolution regime and discuss it in this section, together with the corollaries of this process on the evolution of the disk size.

Fifth, we analyze the importance of the prograde and retrograde orbits in the DM halo, and find that they both contribute to the angular momentum transfer in the system. Interestingly enough, we find that the retrograde DM orbits trapped by the stellar bar can reverse their angular momentum, $J$, and become prograde, being trapped by the stellar bar. They contribute to the fraction of prograde orbits which can resonate with the stellar bar and acts to increase the DM bar strength.  We can observe this by tracking the evolution of a fraction of retrograde orbits (Figure\,\ref{fig:ratio}) and comparing the strength of the DM bar (Figure\,\ref{fig:dma2}).  We elaborate on this interesting process in the next section. 

\subsection{Stellar Bar Growth in Retrograde Models and Dark Matter Orbit Reversals}
\label{sec:growth_reversals}

We start the discussion by addressing the growth of stellar bars during the secular phase of evolution in the retrograde DM halos. At the face value, such a growth is surprising. Along the negative $\lambda$ sequence, the fraction of the prograde DM orbits is decreasing, based on the initial conditions. Such a decrease should reduce the efficiency of angular momentum transfer between the disk and the halo because less DM orbits can resonate with the stellar bar. However, we do not detect this trend which should show up during the secular evolution of stellar bars. What is the reason?

\citet{coll18b} found that the fraction of angular momentum moved by the resonances in each \textit{prograde} model was on par with the fraction of prograde orbits found in the DM halo. P90 has 88\% of orbits rotating with the disk. About 88\% of $J$ lost by the disk in this model was transferred by the resonances. 

This is not the case with the retrograde models. For example, the R90 model which started with only 12\% of DM orbits rotating with the disk, exhibited 53\% of $J$ loss by the disk, which was moved through the resonances at the time of application of the spectral analysis. In all of our models, prograde and retrograde, the stellar bar moves at least $\sim 40\%$ of $J$ by means of the resonant angular momentum transfer. 

In section\,\ref{sec:reversals}, we found that some of the DM orbits, which initially are counter-rotating against the bar tumbling, exchange their angular momentum with the stellar bar, gain $J_\textrm{z}$ and reverse their direction of rotation. Figures\,\ref{fig:reversal1} and \ref{fig:reversalJ} display two examples of such reversals and trapping by the stellar bar. The angular momentum of these orbits, which are negative originally, became positive with the trapping. 

Moreover, Figure\,\ref{fig:ratio2} provides a quantitative measure of importance of this process in the counter-rotating model R90. Though R90 halo is comprised of a majority of retrograde orbits, the disk finds a way to resonate with this DM halo by trapping low-$J$ retrograde DM halo orbits and turning them into prograde orbits precessing in the opposite direction to the original one. They are converted into the low-$J$ prograde DM halo orbits that remain trapped in the bar. This accounts for the increase in the DM bar strength in R90 in Figure\,\ref{fig:dma2} at later times.   

We now wish to confirm that the DM reversals take place indeed when the orbit is trapped by the stellar bar. Figure\,\ref{fig:ratio} shows the radial profiles of the prograde-to-retrograde DM orbit ratios, $\beta(R)$, in the inner halos, i.e., $R < 20$\,kpc and $|z| < 10$\,kpc. The profiles, $\beta(R)$, have been calculated at $t = 0$ and $t = 10$\,Gyr. We observe that initially $\beta$ is flat with $R$ by construction, and at the end of the runs it is peaked at smaller $R$. 

For larger retrograde $\lambda$, $\beta$ increases, mostly in the central region, $R < 10$\,kpc. Here it even reaches unity, which corresponds to the P00 model with $\lambda = 0$. This is a substantial modification with respect to the initial conditions.

This increase in the number of prograde orbits can be noticed in the second row of Figure\,\ref{fig:jdot} as well, where the prograde halo orbits gain more angular momentum for larger retrograde $\lambda$, creating a second local maximum in $\dot J$ after $t\sim 6$\,Gyr. The stellar bar is able to resonate with an increased number of prograde orbits and trap them. Indirectly, this is confirmed by the behavior of the ILR in retrograde halos, where this resonance becomes very prominent (Figure\,\ref{fig:spectra}). We return to this point later on with Figure\,\ref{fig:deltaJ} --- in the R90 halo, the ILR is the most important resonance, with the largest value of $\Delta J$, and is positioned deep inside the bar. In the disk, the CR and OLR becoming less important and instead of moving $J$ to the outer disk, the ILR is preferentially moving $J$ to the DM halo by reversing orbits.

\begin{figure}
\centerline{
\includegraphics[width=0.5\textwidth,angle=0] {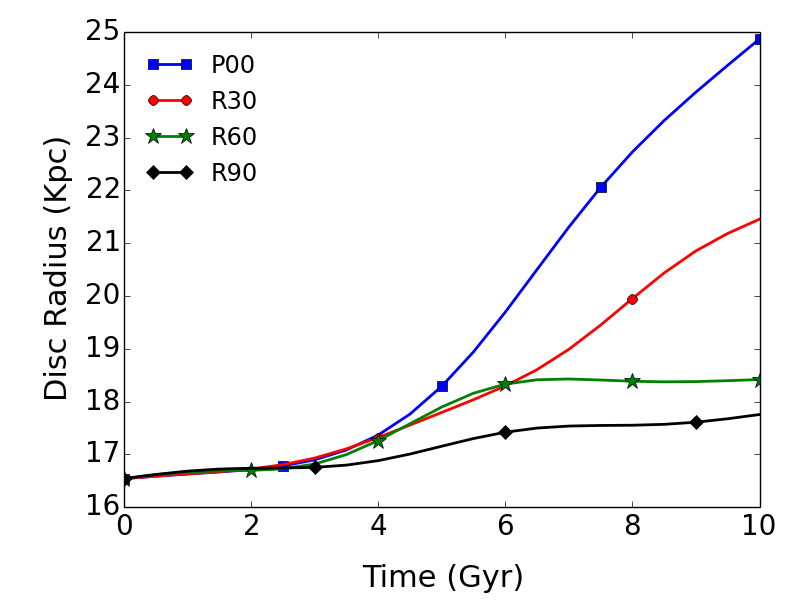}}
\caption{Evolution of disk radius containing $97\%$ of the stellar disk mass for each model. The P00 disk grows substantially which is associated with expanding spiral arms in the outer disk. Contrary to this, the disks within the retrograde halos grow progressively slower and and have smaller sizes by the run end, despite that their bars continue to grow until the end of the simulations, as seen in Figure\,\ref{fig:stellar}a.}
\label{fig:disk_radius}
\end{figure}

\begin{figure*}
\centerline{
\includegraphics[width=\textwidth,angle=0] {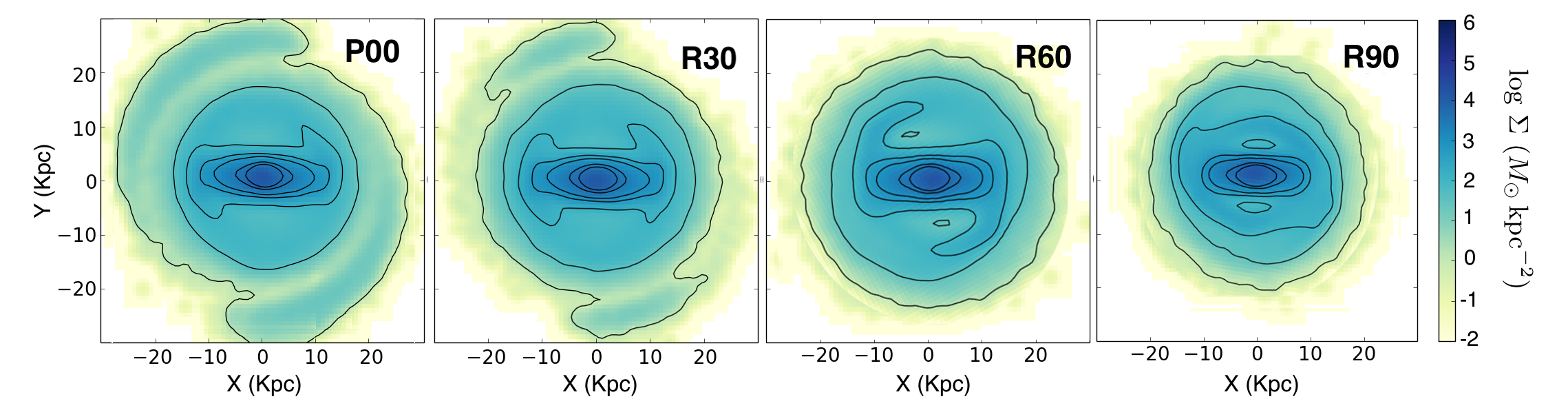}}
\caption{Surface density of stellar disks in retrograde halos at the end of the runs, $t = 10$\,Gyr. The P00 model is shown for comparison. The color palette is identical for all models The contours are linear, with the same limits for each model. Moving along the increasingly negative  $\lambda$, the disks become more compact, and the spiral arms become less obvious, but the bars remain strong and of similar size. The color palette on the right is in units of $M_\odot\,\textrm{kpc}^{-2}$. The stellar disk rotation is anti-clockwise and the retrograde halo spinning in the prograde direction in all models.}
\label{fig:disk_morphology}
\end{figure*}

Together with the Figure\,\ref{fig:ratio2}, which shows the ratio of prograde-to-retrograde orbits, $\beta(t)$ in the same region of the inner halo, as a function of time, Figure\,\ref{fig:jdot} reveals that the rate of angular momentum absorption by the halo peaks exactly when the stellar bar amplitude, $A_2$, accelerates its growth (see Figure\,\ref{fig:stellar}a). It confirms that the majority of reversals occurred after buckling, and that the number of reversals increases with retrograde $\lambda$. Both Figures confirm that reversals compensate for the initially smaller fraction of prograde orbits in retrograde halos. In principle, the increased number of reversals can allow the stellar bar to maintain it's constant trapping efficiency in retrograde halos.

Note, that the increase in the fraction of the prograde DM orbits due to the reversals has little effect on the value of $\lambda$, which is a global property of DM halos. For example, in R90, we observe the largest number of orbit reversals and the largest gain of $J$ over 10\,Gyr for any model. But its retrograde $\lambda$ decreases only from $\lambda=0.0903$ at $t=0$\,Gyr to $\lambda=0.0902$ at $t=10$\,Gyr, which is negligible.

We can summarize that models with retrograde halos do not behave similarly to models with rigid unresponsive DM
halos. We point out two main differences between the unresponsive and live halos. First, along the retrograde
$\lambda$ sequence, the ratio of prograde to retrograde orbits is not negligible. It varies between 1.0 in the
$\lambda=0$ model and 0.12 in the $\lambda=0.09$ model at $t=0$. Second, the orbit reversals in DM halos act to increase the ratio of prograde ot retrograde orbits, and this effect is much more pronounced in the central region which contains the inner resonance, the ILR. The CR and the UHR can be affected as well, but to a lesser degree.

\subsection{Angular Momentum Redistribution and the Disk Size}
\label{sec:disk_size}

Next, we look into corollaries of the angular momentum redistribution in the models, and most importantly how this affects the disk size evolution. Some relation between the angular momentum transfer to the outer disk and the disk size is expected because the stellar orbits in the outer disk are nearly circular, and the only way they can absorb more angular momentum is by increasing their radii, i.e., by disk expansion in the radial direction. The only possible alternative to this evolution is if the angular momentum is directed to the DM halo rather than to the outer disk. If the spiral arms are excited in the outer disk, this is a clear indication that at least some of the angular momentum is absorbed by the region.     

Figure\,\ref{fig:disk_radius} displays the evolution of radii which encompass 97\% of the stellar disk masses.  Such a measurement is comparable to an observational measurement of the 25th magnitude isophote of the galactic disk, $R_{25}$, the Holmberg radius. We observe that a clear hierarchy in disk sizes has developed after the buckling, with the radius of P00 disk growing linearly with time, the R30 disk grows slower, while all other retrograde models nearly saturate after buckling, i.e., $t\sim 6$\,Gyr. Note that the original size of the disk which contains 97\% of its mass is 16.5\,kpc for all prograde and retrograde models. Hence, models with increasingly retrograde $\lambda$ exhibit progressively smaller sizes at the end of the simulations.

We found the same trend along the prograde $\lambda$ sequence --- disk size decreases with increasing $\lambda$. This can be seen, for example, in the Figure\,6 of \citet{coll18a}. In this case, the reason for this behavior is directly related to the stellar bar evolution. Bars within faster spinning prograde halos decay after the buckling. At the end of the runs, the higher $\lambda$ models have substantially lower amplitude, and so less angular momentum is transferred from the underlying disks to the host halos. 

But this explanation does not apply for the retrograde models, because the stellar bars remain strong at the end of the run, although differ in strength by about 20\% among themselves. The spread in the stellar bar sizes by the end of the runs is also about 20\%, with the P00 bar being the longest and R90 the shortest, e.g., Figure\,\ref{fig:stellar}b.  

Hence, it appears that the most substantial growth of the stellar disk occurs in P00 model, and decreases with $\lambda$ along the prograde and retrograde sequences, resulting in measurably smaller disks. These results agree well with the orbital spectral analysis, e.g., Figure\,\ref{fig:spectra}. Fewer stellar orbits are trapped at the OLR with increasing retrograde $\lambda$, because this resonance wanders close to the disk edge as the stellar bar brakes. Little mass resides in this region which can absorb the angular momentum. What is the reason for such an evolution of stellar disks in {\it retrograde} spinning halos? We shall tackle this issue now.

\begin{figure*}
\centerline{
\includegraphics[width=1.15\textwidth,angle=0] {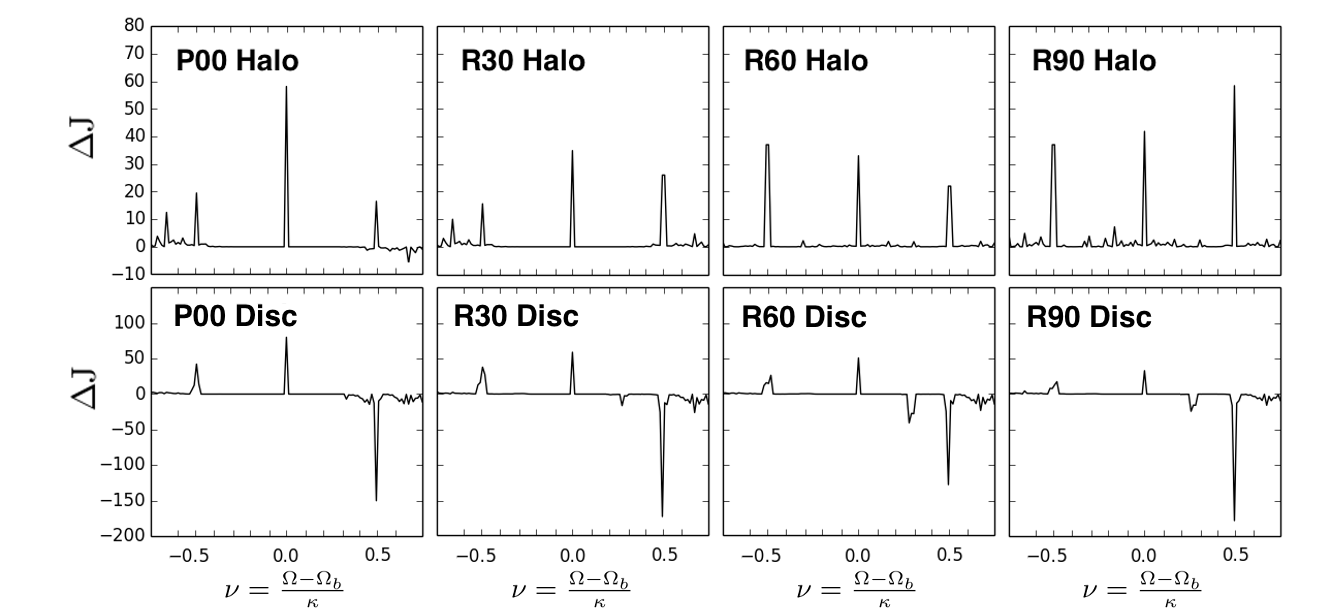}}
\caption{The orbital spectral analysis for the counter-rotating $\lambda$ sequence showing the associated loss of angular momentum, $\Delta J$, by the disk, between $t=8$\,Gyr and 9\,Gyr (bottom), and the same gain $\Delta J$ by the halo (top). The $x$-axis gives the normalized frequency, $\nu\equiv (\Omega-\Omega_{b})/\kappa$ (see definition in the text). The $y$-axis displays the angular momentum gain/loss at each halo and disk resonance, the ILR ( $\nu=0.5$), the CR ($\nu=0.0$), the OLR ($\nu=-0.5$) and the UHR ($\nu=0.25$). The frequencies are binned in $\Delta\nu=0.01$ and $\Delta J$ is given in units of $M_\odot\,\textrm{km}\,\textrm{s}^{-1}\,\textrm{kpc}^{-1}$.
}
\label{fig:deltaJ}
\end{figure*}

\begin{figure}
\centerline{
\includegraphics[width=0.5\textwidth,angle=0] {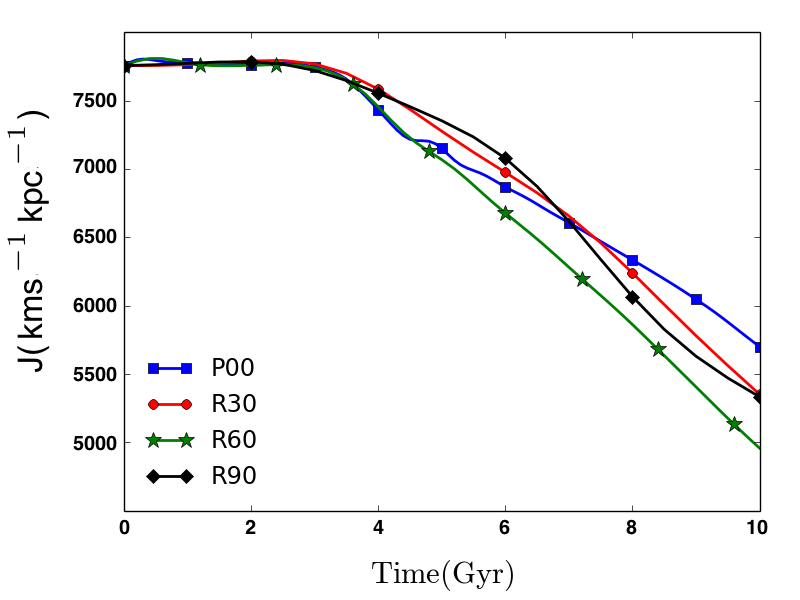}}
\caption{Evolution of the total angular momentum, $J$, in stellar disks. $J$ is in units of
$M_\odot$\,$\textrm{km}\,\textrm{s}^{-1}\,\textrm{kpc}^{-1}$. Compare to Figure\,9 in \citet{saha13} which shows a
similar result, but only before the buckling and only for $\lambda=0.05$ model.}
\label{fig:angmo_1d}
\end{figure}

First, we look for an indication that the spiral activity in the outer stellar disks indeed confirms our claim that in its presence the outer disk absorbs $J$ from the bar region and drives the material out, thus increasing the disk size. Figure\,\ref{fig:disk_morphology} exhibits the surface density of the disks at the end of the runs in retrograde models. In models R60 and R90, we observe only weak outer spirals after the buckling. Those decay completely by the end of the runs. On the other hand, the P00 model displays an ongoing activity in the spiral arms which increases the disk size beyond 25\,kpc. R30 behaves similarly, but the spiral arm generation is less vigorous, and the disk increases less than in P00.

The spiral arm activity of the outer disk can be verified by the rate of angular momentum transfer and by spectral orbit analysis. In the bottom row of Figure\,\ref{fig:jdot}, the angular momentum transfer is not visible at larger radii in the disk, for larger retrograde $\lambda$. The OLR in the P00 disk extends to nearly $R\sim 25$\,kpc, while there is no action of angular momentum transfer in the R90 disk beyond $R\sim 20$\,kpc. This confirms the overall picture discussed above that the outer disk does not absorb $J$ in the high $\lambda$ of retrograde halos emitted from the bar region.

The orbital spectral analysis performed in section\,\ref{sec:orbital_analysis} displays the decreasing efficiency of {\it
stellar} orbit trapping by the outer resonance, the OLR, and associated declining trapping of DM orbits by the main halo
resonance, the CR (for the $\lambda=0$ model) (Figure\,\ref{fig:spectra}). But note the increased trapping efficiency of DM orbits by the halo's ILR and  OLR resonances. We have calculated the angular momentum, $\Delta J$, absorbed and emitted by all the disk and halo main resonances during secular evolution in Figure\,\ref{fig:deltaJ}. 

The main loss of $J$ in the disk is by the ILR. This resonance is strongest in R90, even compared to P00 model. Additional
resonance appears in the disk is the UHR, which also emits $J$. The disk CR resonance exhibits absorption of $J$, but its $\Delta J$ decreases along the retrograde $\lambda$ sequence. Importantly, the disk OLR shows very little absorption of $J$ and looks completely insignificant in R90. 

The main absorption of $J$ in the DM halos switches from the CR in P00 to ILR in R90. In prograde models, we have observed the increase absorption by the halo ILR as well, but all prograde models have been dominated by absorption of $J$ by the CR (see Figure\,11 in \citet{coll18b}). So, the dominant role of the halo ILR resonance is a clear signature of the faster spinning retrograde halo.
 
We plot the 1-D overall $J$ loss with time by the stellar disks (Figure\,\ref{fig:angmo_1d}), which is simpler than the 2-D map in Figure\,\ref{fig:jdot}. It provides complementary information which helps to distinguish both similarities and differences in the total $J$ lost by the disks. To calculate the fraction of angular momentum lost by the resonant and nonresonant interactions in the disk-halo system in our models, we use the orbital spectral analysis shown in Figure\,\ref{fig:deltaJ}, and measure $\Delta J$ lost by each of the resonances.  

We can now quantify the flow of angular momentum in our retrograde models and compare it with P00. We group the resonances into outer ones, which include the CR and the OLR, and the inner one, the ILR, comparing directly the two extreme models, P00 and R90. The amount of angular momentum absorbed by the disk CR and OLR in P00 is $\sim
91\,M_\odot\,\textrm{km}\,\textrm{s}^{-1}\,\textrm{kpc}^{-1}$. The total amount $\Delta J$ lost by the disk during the same time is $\sim 250\,M_\odot\,\textrm{km}\,\textrm{s}^{-1}\,\textrm{kpc}^{-1}$. Hence, about 36\% of this $\Delta J$ has been absorbed by the disk OLR and CR.

On the other hand, $\Delta J$ absorbed by the disk CR and OLR in R90 is $\sim
50\,M_\odot\,\textrm{km}\,\textrm{s}^{-1}\,\textrm{kpc}^{-1}$. The total amount $\Delta J$ lost by the disk during the same time is $\sim 725\,M_\odot\,\textrm{km}\,\textrm{s}^{-1}\,\textrm{kpc}^{-1}$. This means that only $\sim 7\%$ of this $\Delta J$ has been absorbed by the disk OLR and CR. This is about five times less than in P00, and appears to be the reason why the disk in R90 expand much less than in P00.

To summarize, the disk size growth depends on the fraction of angular momentum which is transferred to the outer disk, in the retrograde models. We find that the stellar disks inside the faster retrograde halos preferentially communicate with the inner halo rather than with the outer disk.  The lack of the outer spiral arms is observable, and a clear imprint of the parent retrograde halo.

\section{Conclusions}
\label{sec:concl}

We have performed a high-resolution study of stellar bars and associated DM response in \textit{retrograde} spinning DM halos, with $\lambda \sim 0-0.09$. We find that evolution of stellar bars in these halos differs substantially from that in the prograde halos studied in \citet{coll18a,coll18b}. Moreover it differs from evolution of stellar bars in the frozen unresponsive halos. The DM response to the stellar bar perturbation in retrograde halos is modified from their prograde counterparts. We have analyzed the orbital structure and the angular momentum transfer in these disk-halo systems, and focused on the role of the resonances in transferring the angular momentum. 

While statistics of retrograde galactic DM halos or their components is not available at present, numerical and theoretical modeling point to their possibility, and observations reveals individual cases of counter-rotating components in the DM and stars. We summarize our results below.

First, we find that the bar instability is slowed down with increasingly counter-rotating $\lambda$ --- a trend first noticed by \citet{saha13} for a single model with $\lambda=-0.05$. Together with the prograde models, the retrograde models form a monotonic sequence of prolonging the characteristic timescale of bar instability.

Second, and probably the most dramatic difference between the prograde and retrograde models, lies in the secular evolution of the stellar bars. While increasing prograde $\lambda$ ultimately damps the stellar bars, which basically dissolve leaving a weak oval distortion, increasing the retrograde halo spin has only a minor effect on the secular evolution of the stellar bars. Their amplitudes and sizes differ by $\sim 20\%$ at the end of the runs. We have quantified the rate of the angular momentum transfer from the stellar disk to the DM halo, and analyzed the role of the prograde and retrograde orbits in this process.

Third, the DM response to the stellar bar is much weaker in all retrograde models, substantially weaker than in the prograde ones --- the $m=2$ mode Fourier amplitudes for this DM response is a factor of a few weaker during the bar instability, the dynamical phase of the evolution. Moreover, while in prograde models the DM response is aligned with stellar bar, in retrograde models it is nearly orthogonal to the bar. This emphasizes the increased importance of the ILR in the $J$-transfer for the latter models. 

The DM response to the stellar bar perturbation during the secular phase includes trapping of DM orbits and angular momentum transfer by the main resonances. We find that both the trapping of DM orbits and $J$ transfer from the disk to the halo at this stage do not depend on the retrograde spin $\lambda$, in stark difference with the prograde models. Yet, the contribution of individual resonances does change along the retrograde sequence, as the orbital spectral analysis shows.  The efficiency of the resonance trapping remains at $\sim 20\%$ of the sampled orbits. This percentage includes the main resonances, the ILR, CR and the OLR.  In the nonrotating P00 halo model, the most important resonance for the angular momentum absorption by the DM halo is the CR followed by the OLR \citep{atha03,marti06,dubi09}. Our results show that the ILR resonance replaces the CR resonance in trapping the DM orbits and angular momentum absorption in retrograde halos, which has not been seen before.

Finally, the strength of the DM response and DM orbit trapping by the resonances during  the secular phase of evolution is regulated by two factors. One is trivially related to the initial conditions --- the fraction of prograde DM orbits decreases with $\lambda$ by construction. Another one is much more interesting --- ability of trapped DM orbits to reverse their angular momentum. These reversals have modified substantially the fraction of prograde DM orbits, and this change is heavily weighted towards the central regions of about few kpc. Not surprisingly, the rversal are dominated by the low-$J$ DM orbits. The fraction of prograde DM orbits in this regions has increased to about unity. Both the streaming along the orbits and its precession has been reversed. This process has never been discussed in the context of disk-halo interactions. Reversals of DM orbits have been responsible for increased trapping and angular momentum losss by the inner disk, thus resulting in increased strength of stellar bars during secular evolution phase.  
 
Reversals of DM orbits have led to a number of corollaries, of which we point out the most important one --- evolution of
the disk size. We find that stellar disks with low retrograde $\lambda$ pump a substantial fraction of angular momentum
into the outer disk via the OLR. This leads to expansion of the outer disk and ongoing spiral activity there. However, with
increasing $\lambda$, the fraction of prograde stellar orbits in the outer disk decreases, by construction. The angular
momentum transfer to the outer disk decreases by a factor of $\sim 5$ from $\lambda=0$ to $\lambda=-0.09$. Instead, this
angular momentum is diverted to the inner halo, which absorbs it mostly via the ILR, and by the CR to a lesser degree.

Stellar disks in these halos do not show spiral arm activity in the outer region and do not grow with time. When comparing
the disks sizes, both prograde and retrograde models of larger $\lambda$ show smaller disks. In fact, the P00 model with
$\lambda=0$ has the largest disk and most prominent spiral arms at the end of the simulation. In prograde models this is
explained by the bar dissolution after buckling which prevents the movement of angular momentum to the outer disk. In
the retrograde models, it follows from reduced activity of the disk OLR and CR --- the disk does not expand, being much
less efficient in absorbing the angular momentum by its outermost part. The angular momentum is absorbed instead by the increased efficiency of the halo OLR.

In summary, as stellar bars are the prime \textit{internal} factor which drives the galaxy evolution, they warrant a very careful numerical study. The steady interest in the stellar bar evolution over the last few decades is due to their ability to link the DM halo  to the stellar disk through angular momentum transfer. Studying dynamics of the disk-halo systems consistently reveals new effects which are important for galaxy evolution. Through careful study, we begin to understand the observational corollaries of stellar bars action and its affect on the dynamics of the inner DM halo.

\section*{Acknowledgements}
We acknowledge helpful discussions with Scott Tremaine and Alar Toomre, and are grateful to Jorge Villa-Vargas and Emilio Romano-Di\'az for the help with numerical issues. This work has been partially supported by the HST/STScI Theory grant AR-14584, and by JSPS KAKENHI grant \#16H02163 (to I.S.). I.S. is grateful for support from International Joint Research Promotion Program at Osaka University. The STScI is operated by the AURA, Inc., under NASA contract NAS5-26555.   Simulations have been performed on the University of Kentucky DLX Cluster.


\end{document}